\documentclass[epj]{svjour}
\usepackage{graphics}
\begin{document}
\title{Interaction of phonons with discrete breather in strained graphene}
\author{Iman~Evazzade\inst{1} \and Mahmood~Rezaee~Roknabadi \inst{1}
\thanks{\emph{Corresponding author:} roknabad@um.ac.ir} \and Mohammad~Behdani\inst{1} \and Fatemeh~Moosavi \inst{2} \and Daxing~Xoing \inst{3} \and Kun~Zhou \inst{4} \and Sergey~V.~Dmitriev \inst{5,6} % etc
}                     % Do not remove
\institute{Department of Physics, Faculty of Science, Ferdowsi University of Mashhad, Mashhad, Iran \and Department of Chemistry, Faculty of Science, Ferdowsi University of Mashhad, Mashhad, Iran \and Department of Physics, Fuzhou University, Fuzhou 350108, Fujian, China \and School of Mechanical and Aerospace Engineering, Nanyang Technological University, 50 Nanyang Avenue, Singapore 639798, Singapore \and Institute for Metals Superplasticity Problems, Russian Academy of Sciences, 450001 Ufa, Russia \and National Research Tomsk State University, Lenin Ave, 36, 634050 Tomsk, Russia}
\date{Received: date / Revised version: date}
\abstract{
We numerically analyze the interaction of small-amplitude phonon waves with standing gap discrete breather (DB) in strained graphene. To make the system support gap DB, strain is applied to create a gap in the phonon spectrum. We only focus on the in-plane phonons and DB, so the issue is investigated under a quasi-one-dimensional setup. It is found that, for the longitudinal sound waves having frequencies below 6 THz, DB is transparent and thus no radiation of energy from DB takes place; whereas for those sound waves with higher frequencies within the acoustic (optical) phonon band, phonon is mainly transmitted (reflected) by DB, and concomitantly, DB radiates its energy when interacting with phonons. The latter case is supported by the fact that, the sum of the transmitted and reflected phonon energy densities is noticeably higher than that of the incident wave. Our results here may provide insight into energy transport in graphene when the spatially localized nonlinear vibration modes are presented.
 % end of PACS codes
} %end of abstract
\maketitle

\section{Introduction}

Graphene is a two-dimensional, one-atom-thick carbon crystal. Its physical and mechanical properties are promising for a number of nanotechnology applications~\cite{Geim,Novoselov}. For example, graphene has a high melting temperature about 5000~K~\cite{Melting,Melting1}, high velocity of the longitudinal sound waves about 20~km/s~\cite{SoundVel}, and high thermal conductivity~\cite{SKH2010}. Nowadays energy transport in graphene carried by phonons and nonlinear excitations has been an interesting topic due to the rapidly growing field of {\em phononics}~\cite{Phononics,Phononics1}. In this field, the relevant thermal transistors~\cite{Transist1,Transist2}, thermal diodes~\cite{Diod1,Diod3,Isotops,Diod2,Diod4}, and thermal logic gates~\cite{Gates1,Gates2,Gates3} are still being developed. Such developments require a better theoretical and experimental understanding on anomalous thermal transport in low-dimensional systems~\cite{Lepri,Dhar,Anomaly1,Anomaly2,Anomaly5,Anomaly6,Anomaly7,Anomaly8}. Here we also note some progress in harmonic systems \cite{Harm1,Harm2,Harm3,Harm4,Harm5,Harm6}.

Atomistic simulation is a powerful tool in studying linear and nonlinear lattice dynamics of graphene and several theoretical models. With such a method, it has been shown that, the radial coefficient of thermal expansion of graphene nano-scrolls is two orders of magnitude greater than that of diamond~\cite{thermexp1,thermexp2,thermexp3}; randomly distributed defects of graphene can significantly influence energy transport~\cite{SKH2010,Isotops,LeKr2007}; Fourier law can be recovered in nonlinear chains with breakable interatomic bonds (similar to the real crystals)~\cite{SavKos}; phonon localization and thermal rectification can appear in the chains with strain gradient~\cite{Gradient}. Atomic-scale simulation also helps to understand the structural transformations of carbon and hydrocarbon polymorphs~\cite{KM1,KM2,KM3,KM4}. Of particular interest is the interaction between phonons and other nonlinear lattice excitations, termed as either discrete breathers (DB) or intrinsic localized modes (ILM)~\cite{Dolgov,ST1988,FW1998,FW2008,Uspekhi}. Properties of DB in graphene and carbon nanotubes have been studied in a number of works~\cite{DB1,DB2,DB3,DB4,DB5,DB6,DB7,DB8,DB9,DB10,DB11,DB12,DB13,DB14,DB15,DB16,DB17,DB18,DB19}. In theoretical models, DB~\cite{XZ2016,XSD2017,pDB1,pDB2} and solitons~\cite{JYZZ2017} have been shown to affect thermal transport in certain nonlinear chains. In a recent study~\cite{Evazzade}, the supratransmission effect has been observed for a harmonically driven strained graphene nanoribbon, even at small driving amplitudes~\cite{Supra1,Supra2,Supra3}. In such cases, energy is transported not by running DB, as in some classical works~\cite{Supra1,Supra2,Supra3}, but by phonons emitted by standing DB with time-modulated amplitude excited next to the driven carbon atoms~\cite{Saadatmand}.

Motivated by such a background, we address the issue of phonon scattered by gap DB in a strained graphene. We are interesting to explore how DB interacts with small-amplitude phonons of different wavelengths, and in turn how phonons affect DB. Such information might help to understand the effect of DB on energy transport in strained graphenes.

\begin{figure}
\resizebox{0.5\textwidth}{!}{%
	\includegraphics{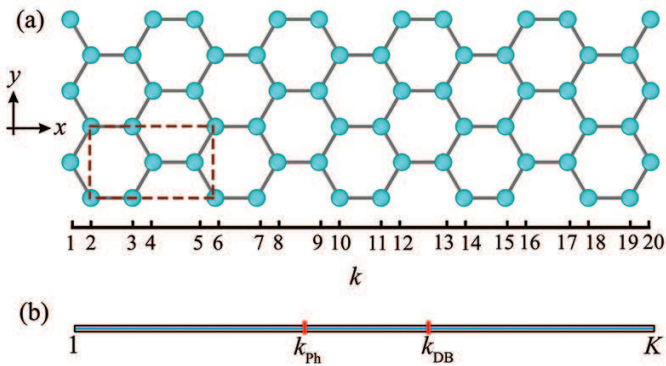}
}
\caption{(a) Graphene nanoribbon with armchair (zigzag) directions along the $x$ ($y$) axis. Atomic rows parallel to the $y$ axis are numbered by index $k$. Rectangular translational cell containing four atoms is shown by dashed lines. In simulations 100 translational cells are used in the $x$ direction and one cell in the $y$ direction. (b) Scheme showing location of the phonon source at $k_{\rm ph}=160,\,161$ and DB at $k_{\rm DB}=240,\,241$ in the computational cell of $K=400$ atomic rows.}
\label{fig:1}
\end{figure}

\section{Simulation setup}
\label{SimulationSetup}
In Fig.~\ref{fig:1}(a), the initial structure of graphene nanoribbon is schematically plotted, together with the Cartesian coordinates with the $x$ ($y$) axis along the armchair (zigzag) direction. The dashed lines show the rectangular translational cell with four carbon atoms. A rectangular supercell with 100 translational cells along the $x$ axis and just one cell along the $y$ axis is used in simulations. Atomic rows parallel to the $y$ axis are numbered with the index $k$. Periodic boundary conditions are employed along both the $x$ and $y$ axes.

A standard set of interatomic potentials developed by Savin et al~\cite{SKH2010} are used to describe interatomic interactions. It reproduces the dispersion curves for graphene better than that by Brenner potentials~\cite{brennerPot1990}.

\begin{figure}
\resizebox{0.5\textwidth}{!}{%
	\includegraphics{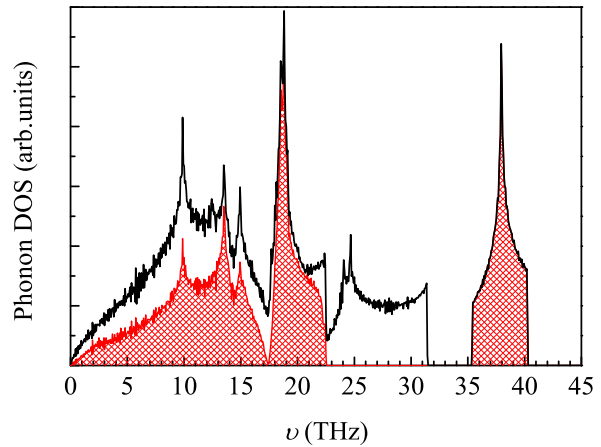}
}
\caption{Phonon density of states for graphene homogeneously strained with $\varepsilon_{xx}=-0.1$, $\varepsilon_{yy}=0.35$, and $\varepsilon_{xy}=0.0$. Shaded DOS represents the in-plane phonons, while blank DOS stands for phonons with atomic vibrations normal to the graphene sheet.}
\label{fig:2}
\end{figure}

Applying in-plane elastic strain with components $\varepsilon_{xx}=-0.1$, $\varepsilon_{yy}=0.35$, and $\varepsilon_{xy}=0.0$ introduces a gap into the phonon spectrum of graphene. In Fig. \ref{fig:2}, the phonon density of states (DOS) of this strained graphene is shown. Shaded (blank) DOS stands for the in-plane phonons (phonons with atomic displacements normal to the $(x,y)$ plane). As mentioned, we only focus on the in-plane motion of atoms, so that only shaded DOS is of importance here.

Small-amplitude in-plane phonons propagating along the $x$ axis are excited by ac driving of two atomic rows with $k_{\rm ph}=160,\,161$. The location of this phonon source is schematically shown in Fig.~\ref{fig:1}(b). The driven atomic rows are forced to move along the $x$ axis according to the harmonic law 
\begin{equation}\label{Driving}
\Delta x_{160}(t)=-\Delta x_{161}(t)=\varepsilon \sin(2\pi \nu t),
\end{equation}
where the driving amplitude is set to be small, i.e., $\varepsilon=10^{-3}$\AA$\,$ and the driving frequency $\nu$ is within the range of shaded DOS as shown in Fig.~\ref{fig:2}. A large-amplitude DB is excited by giving its initial displacements as follows: $\Delta x_{k}(0)=- 0.1875$\AA$\,$ for $k<240$; $\Delta x_{k}(0)=- 0.45$\AA$\,$ for $k=240$; $\Delta x_{k}(0)= \,\,\,\, 0.45$\AA$\,$ for $k=241$, and $\Delta x_{k}(0)= \,\,\,\, 0.1875$\AA$\,$ for $k>241$. Fig.~\ref{fig:1}(b) also shows the location of DB as $k_{\rm DB}$. In addition, initially all atoms are set as zero velocities, i.e., $\Delta\dot{x}_k(0)=0$ for $k=1,...,400$, and they are shifted away from the excited DB by $0.1875$\AA$\,$ to consider possible local lattice expansion induced by the DB.

Such an initial setup results in a DB which emits parts of its energy in the form of small-amplitude radiations during a transient period. These additional radiations are absorbed, and after a stable DB has been obtained, the phonon source then is turned on. The source emits phonons to the left and right with the amplitude and frequency equal to the driving amplitude $\varepsilon$ and driving frequency $\nu$, respectively. The duration of numerical run is equal to $T=2S/v_g$, where $S$ is the distance from the phonon source to DB [80 translational cells shown in Fig.~\ref{fig:1}(a)] and $v_g$ is the group velocity of the excited phonon. So, the time for the phonon to reach DB at $k=k_{\rm DB}$ is $T/2$, and after time $T$, the phonon will be reflected by DB and come back to the source initially at $k=k_{\rm ph}$. On the other hand, the phonon emitted to another direction will reach the end of the computational cell, $k=1$, in time $T$. In addition, a part of phonon emitted to the right from the source can go through the DB and by time $T$ it will propagate 80 translational cells, i.e., from $k=k_{\rm DB}=240$ to $k=320$.

Bearing the above picture, at time $T$ the energy per atom can be calculated by considering the following three regions:
\begin{eqnarray}\label{eI}
e_{\rm I}=\frac{1}{159}\sum_{k=1}^{159}e_k, 
\end{eqnarray}
\begin{eqnarray}\label{eR}
e_{\rm R}=\frac{1}{74}\sum_{k=162}^{235}e_k-e_{\rm I}, 
\end{eqnarray}
\begin{eqnarray}\label{eT}
e_{\rm T}=\frac{1}{74}\sum_{k=246}^{319}e_k, 
\end{eqnarray}
where $e_k$ is the total (kinetic plus potential) energy of $k$-th atom. Energy per atom $e_{\rm I}$ gives the energy density of the incident phonon, while $e_{\rm R}$ and $e_{\rm T}$ provide the energy densities carried by the phonon reflected and transmitted by the DB, respectively. In practice, there are certain additional energies emitted by the DB. To describe this, we define the energy of DB as
\begin{eqnarray}\label{EDB}
E_{\rm DB}=\sum_{k=236}^{245}e_k. 
\end{eqnarray}

\begin{figure}
\resizebox{0.45\textwidth}{!}{%
	\includegraphics{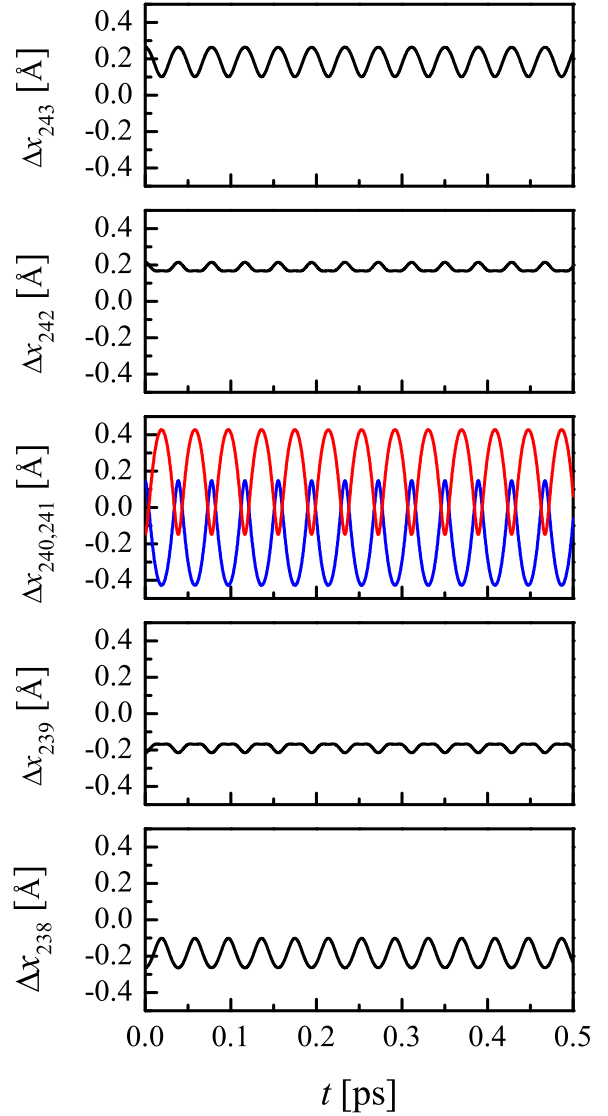}
}
\caption{Displacements of atoms in the vicinity of DB as a function of time. DB is centered at the covalent bond connecting atoms $k=240$ and $k=241$.}
\label{fig:3}
\end{figure}

\section{Numerical results}
\label{Numerics}
In Fig.~\ref{fig:3} the displacements of atoms in the vicinity of DB are shown as a function of $t$. From this, one is able to measure DB’s frequency $\nu_{\rm DB}=25.67$ THz and DB’s amplitude $A_{\rm DB}=0.1438$ \AA. Clearly, DB’s frequency lies in the gap of shaded DOS for the in-plane phonons as shown in Fig.~\ref{fig:2}. DB is mainly located at six atoms from $k=238$ to $k=243$ since the vibration amplitudes of other atoms are very small. So it is reasonable that in calculation of DB’s energy, we only consider 10 atoms as formulated in Eq.~(\ref{EDB}). One should note that the standing DB, which is not disturbed by phonons, has an extremely long lifetime, as it practically does not radiate any energy.

\begin{figure}
\resizebox{0.5\textwidth}{!}{%
	\includegraphics{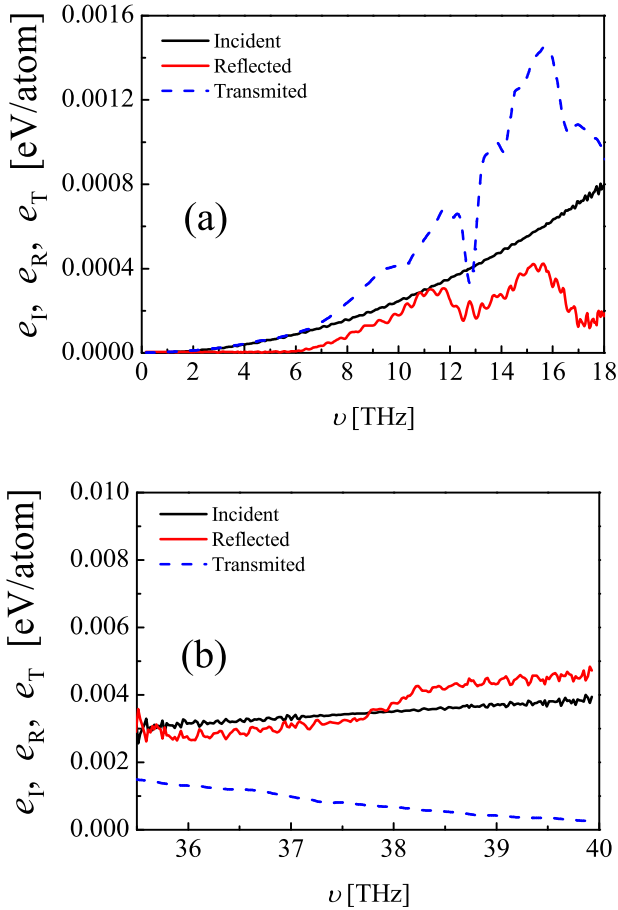}
}
\caption{Energy densities $e_{\rm I}$, $e_{\rm R}$, and $e_{\rm T}$ as a function of driving frequency, calculated at the end of numerical run by Eqs.~(\ref{eI}-\ref{eT}), respectively: (a) for acoustic and (b) for optical phonons.}
\label{fig:4}
\end{figure}

\begin{figure*}
\resizebox{0.95\textwidth}{!}{%
\includegraphics{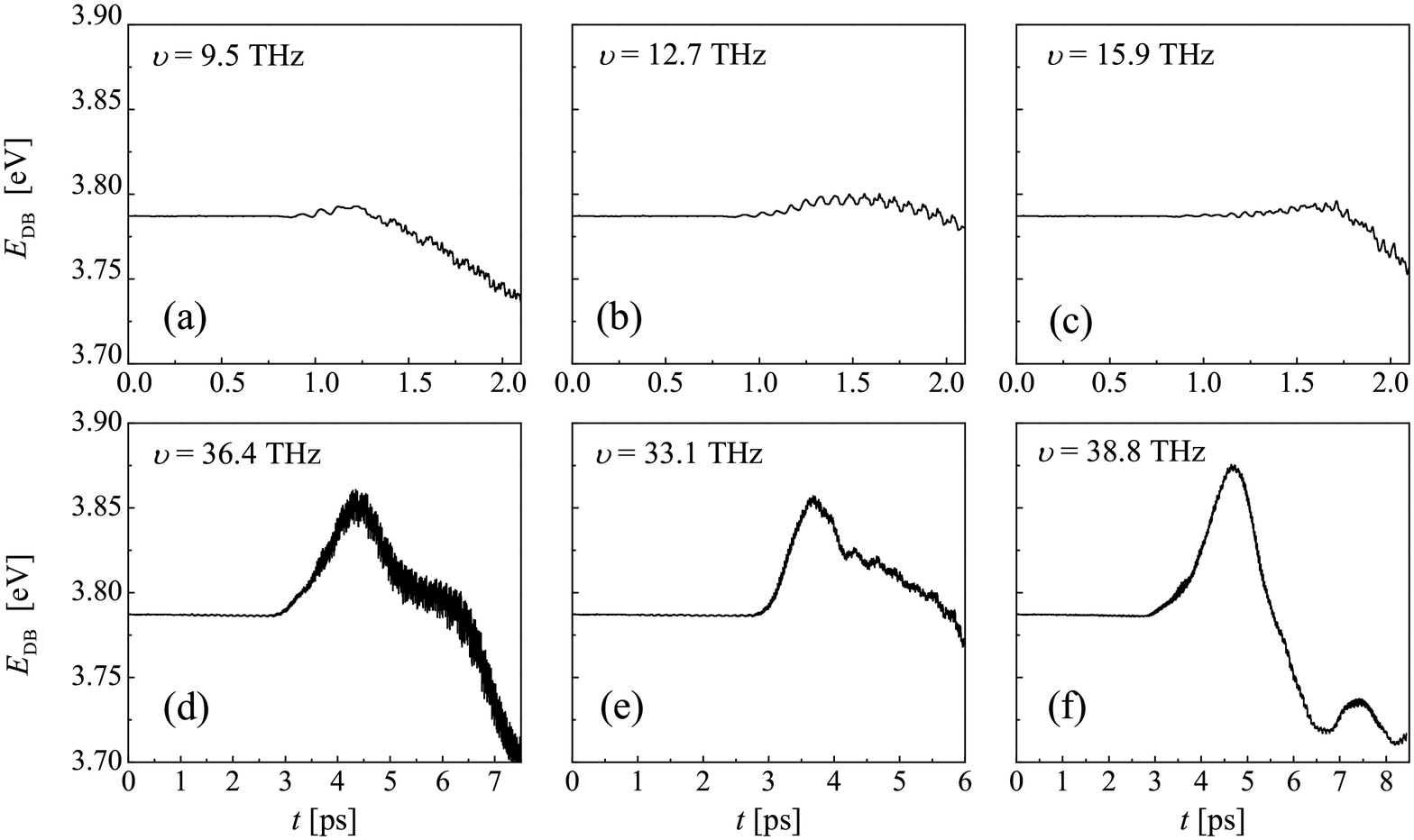}
}\caption{DB's energy as a function of time. The driving frequency is indicated in each panel.}
\label{fig:5}
\end{figure*}

We next turn to the result of energy densities $e_{\rm I}$, $e_{\rm R}$, and $e_{\rm T}$, calculated at the end of numerical run by Eqs.~(\ref{eI}-\ref{eT}), respectively. This result is presented in Fig.~\ref{fig:4} as a function of driving frequency within the acoustic [see Fig.~\ref{fig:4}(a)] and optical [see Fig.~\ref{fig:4}(b)] phonon bands. 

From Fig.~\ref{fig:4}(a), as to the acoustic case, for the driving frequencies $\nu<6$~THz, the energy density of reflected phonon, $e_{\rm R}$, is negligible, so the incident and transmitted densities ($e_{\rm I}$ and $e_{\rm T}$) are nearly equal. This indicates that DB is transparent for the long-wavelength phonons with frequencies less than $6$~THz. Such phonons do not cause any radiation from DB since $e_{\rm I}=e_{\rm R}+e_{\rm T}$. The situation is quite different for the driving frequencies above $6$~THz. At first, we should note that in this frequency domain $e_{\rm I}<e_{\rm R}+e_{\rm T}$. This can only be understood from the radiation of energy by DB when interaction with the incident phonon. Besides, in this case $e_{\rm R}$ is always smaller than $e_{\rm T}$, suggesting that the acoustic phonons always pass through the DB only with a much smaller amount of reflection. 

From Fig.~\ref{fig:4}(b), since the optical phonons are always with high frequencies, we do not see a $e_{\rm I}=e_{\rm R}+e_{\rm T}$ behavior as acoustic case under a frequency threshold. While for the another $e_{\rm I}<e_{\rm R}+e_{\rm T}$ domain, we find that $e_{\rm T}$ is considerably smaller than $e_{\rm R}$, in contrast to the $e_{\rm R}<e_{\rm T}$ case shown for the acoustic phonons.

We finally explain why DB radiates energy when it is exposed to a phonon with a high frequency. In Fig.~\ref{fig:5} we depict the result of DB’s energy as a function of time for different driving frequencies. Different panels correspond to the results of different frequencies. Panels (a-c) give the results acoustic phonons, while in (d-f) the results for optical phonons are presented. 

It can be seen that DB’s energy is constant until phonon emitted by the energy source approaches it. Then an increase of $E_{\rm DB}$ and finally followed by a decrease is observed. The increase happens since the phonon enters the region of ten atoms over which the DB energy is summed, see Eq.~(\ref{EDB}). So, the energy excess is actually due to the additional energy of the phonon. Interestingly, for the phonon frequencies around $\nu=12.7$~THz, $E_{\rm DB}$ nearly does not change [see Fig.~\ref{fig:5}(b)]. This can be indicated from Fig.~\ref{fig:4}(a), where around $\nu=12.7$~THz, a half of the DB’s frequency, there is a clear decrease of  both $e_{\rm R}$ and $e_{\rm T}$, so that the relation $e_{\rm I}=e_{\rm R}+e_{\rm T}$ here nearly holds. This evidence supports that in this case the energy radiation from DB does become weak.

\section{Conclusions}
\label{Conclusion}

Interaction of longitudinal phonons propagating in a strained graphene along the armchair direction with a large-amplitude gap DB has been numerically studied. Some useful evidences have been revealed:
\begin{itemize}
\item Long-wavelength acoustic phonons with frequency below $6$~THz do not interact with DB, in that sense that such phonons penetrate through DB and they do not cause any noticeable radiation of energy by DB;

\item Acoustic phonons with frequency above $6$~THz are mainly transmitted when interact with DB. They cause noticeable radiation from DB in the form of small-amplitude running waves. 

\item Optic phonons are mainly reflected by DB. The radiation of energy of DB in this case is less pronounced in comparison to the acoustic phonons with frequencies above 6~THz.

In short, our results here may shed some light on the energy transport behavior carried by phonons when the large-amplitude spatially localized excitations, such as DB, are presented. While we have to note that here only the quasi-one-dimensional treatment are undertaken, further extended studies might require to consider a more realistic two-dimensional setting, where a plane-wave rather than a chain phonon is scattered with the DB localized on a single covalent bond.

\end{itemize}

\section*{Acknowledgments}
I.E. greatfully acknowledges financial support provided by Ferdowsi University of Masshad, grant no. 3/43318 .Work of D.X. was supported by the National Natural Science Foundation of China (Grant No. 11575046), the Natural Science Foundation of Fujian Province, China (Grant No. 2017J06002), and the Qishan Scholar Research Fund of Fuzhou University, China. Work of S.V.D. was supported by the Russian Science Foundation, grant No. 16-12-10175. This work was supported by the Russian Foundation for Basic Research, grant No. 17-02-00984.

\end{document}